\documentclass[preprint,5p,11pt,twocolumn,sort&compress]{elsarticle}

\usepackage{amssymb,amsmath,colonequals,xspace,tikz}
\usetikzlibrary{calc,decorations.pathmorphing,fit,backgrounds,positioning,arrows,shapes}
\tikzstyle{knoten}=[circle,draw=black,thin,fill=white,inner sep=0pt,minimum size=4.5mm]
\tikzstyle{knotenklein}=[circle,draw=black,thin,fill=white,inner sep=0pt,minimum size=2.5mm]

\usetikzlibrary{fadings}
\usetikzlibrary{intersections}

\global\long\def\NP{\mathcal{NP}}

\newtheorem{theorem}{Theorem}

\newproof{proof}{Proof}
\newdefinition{definition}{Definition}

\def\OO{\mathcal{O}}

\newcommand{\pt}{\ensuremath{\widetilde{p}}}

\newcommand{\floor}[1]{
	\left\lfloor #1 \right\rfloor
}
\newcommand{\ceil}[1]{
	\left\lceil #1 \right\rceil
}

\newcommand{\citets}[1]{\citeauthor{#1}'s (\citeyear{#1})}

\makeatletter
\def\NAT@spacechar{~}
\makeatother

\makeatletter
\def\ps@pprintTitle{%
 \let\@oddhead\@empty
 \let\@evenhead\@empty
 \def\@oddfoot{}%
 \let\@evenfoot\@oddfoot}
\makeatother
\usepackage{mathpazo}
\usepackage[euler-digits,T1]{eulervm}

\begin{document}

\begin{frontmatter}

\title{An FPTAS for the Knapsack Problem with Parametric Weights\tnoteref{titleref}}

\author[TUKL]{Michael Holzhauser\corref{cor1}}
\ead{holzhauser@mathematik.uni-kl.de}

\author[TUKL]{Sven O. Krumke}
\ead{krumke@mathematik.uni-kl.de}

\cortext[cor1]{Corresponding author. Fax: +49 (631) 205-4737. Phone: +49 (631) 205-2511}

\address[TUKL]{\normalfont{University of Kaiserslautern, Department of Mathematics\\
  Paul-Ehrlich-Str.~14, D-67663~Kaiserslautern, Germany}}

\begin{abstract}
	\setlength{\parindent}{0pt}%
	In this paper, we investigate the parametric weight knapsack problem, in which the item weights are affine functions of the form~$w_i(\lambda) = a_i + \lambda \cdot b_i$ for $i \in \{1,\ldots,n\}$ depending on a real-valued parameter~$\lambda$. The aim is to provide a solution for all values of the parameter. It is well-known that any exact algorithm for the problem may need to output an exponential number of knapsack solutions. We present the first fully polynomial-time approximation scheme (FPTAS) for the problem that, for any desired precision~$\varepsilon \in (0,1)$, computes $(1-\varepsilon)$-approximate solutions for all values of the parameter. Our FPTAS is based on two different approaches and achieves a running time of $\OO(n^3/\varepsilon^2 \cdot \min\{ \log^2 P, n^2 \} \cdot \min\{\log M, n \log (n/\varepsilon) / \log(n \log (n/\varepsilon) )\})$ where $P$ is an upper bound on the optimal profit and $M \colonequals \max\{W, n \cdot \max\{a_i,b_i: i \in \{1,\ldots,n\}\}\}$ for a knapsack with capacity~$W$.
\end{abstract}


\end{frontmatter}

\section{Introduction}
\label{sec:Intro}

The knapsack problem is one of the most fundamental combinatorial optimization problems: Given a set of $n$ items with weights and profits and a knapsack capacity, the task is to choose a subset of the items with a maximum profit such that the weight of these items does not exceed the knapsack capacity. The problem is known to be weakly $\NP$-hard and solvable in pseudo-polynomial time. Moreover, several constant factor approximation algorithms and approximation schemes have been developed for the problem \citep{IbarraKimKnapsack,LawlerCombinatorialOptimization,MagazineOguzKnapsack,KellererPferschyFPTAS,KellererPferschyFPTAS2} (cf. \citep{Knapsack} for an overview).

In this paper, we investigate a generalization of the problem in which the weights are no longer constant but affine functions depending on a parameter~$\lambda \in \mathbb{R}$. More precisely, for a knapsack with \emph{capacity}~$W$ and for each \emph{item}~$i$ in the \emph{item set}~$\{1,\ldots,n\}$ with \emph{profit}~$p_i \in \mathbb{N}_{> 0}$, the \emph{weight}~$w_i$ is now of the form $w_i(\lambda) \colonequals a_i + \lambda \cdot b_i$ with $a_i,b_i \in \mathbb{Z}$. The resulting optimization problem can be stated as follows:
\begin{align*}
	p^*(\lambda) = \max\ & \sum_{i=1}^n p_i \cdot x_i \\
	& \sum_{i=1}^n (a_i + \lambda \cdot b_i) \cdot x_i \leq W \\
	& x_i \in \{0,1\} \quad \forall i \in \{1,\ldots,n\}
\end{align*}
The aim of this \emph{parametric weight knapsack problem} is to return a partition of the real line into intervals~$(-\infty,\lambda_1], [\lambda_1,\lambda_2], \ldots, [\lambda_{k-1},\lambda_k], [\lambda_k,+\infty)$ together with a solution~$x^*$ for each interval such that this solution is optimal for all values of $\lambda$ in the interval.

Besides the fact that the parametric weight knapsack problem is clearly $\NP$-hard to solve since it contains the traditional knapsack problem, it was shown that any exact algorithm for the problem may need to return an exponential number of knapsack solutions in general \citep{BurkardPferschyInverseParametricKnapsack}. In this paper, we are interested in a \emph{fully polynomial time approximation scheme} for the parametric weight knapsack problem. We will show that, for any desired precision~$\varepsilon \in (0,1)$, a polynomial number of intervals suffices in order to be able to provide a $(1-\varepsilon)$-approximate solution for each $\lambda \in \mathbb{R}$.

In the following, we let $P \leq \sum_{i=1}^n p_i$ denote an upper bound on the optimal profit and set
\begin{align}\label{eqn:BigM}
	M \colonequals \max\{W, n \cdot \max\{a_i,b_i: i \in \{1,\ldots,n\}\}.
\end{align}

\subsection{Previous work}

A large number of publications investigated parametric versions of well-known problems. This includes the parametric shortest path problem \citep{KarpOrlinParametricShortestPath,YoungTarjanOrlinParametricShortestPath,CarstensenParametricProblemsDisseration,MulmuleyShortestPathLowerBound}, the parametric minimum spanning tree problem \citep{FernandezParametricMinimumSpanningTree,AgarwalParametricMinimumSpanningTree}, the parametric maximum flow problem \citep{GalloParametricMaxFlow,McCormickParametricMaxFlow,ScutellaParametricMaxFlow}, and the parametric minimum cost flow problem \citep{CarstensenParametricProblems} (cf. \citep{SpundParametricKnapsack} for an overview).

A problem that is related to the parametric weight knapsack problem considered here is the parametric \emph{profit} knapsack problem, in which the weights are constant but the profits take on an affine form~$p_i(\lambda) = a_i + \lambda \cdot b_i$. \citet{CarstensenParametricProblems} first showed that the number of breakpoints of the optimal profit function may be exponentially large in general and pseudo-polynomially large in the case of integral input data. \citet{ChaimeParametricKnapsack} later presented a pseudo-polynomial exact algorithm for this problem. \citet{SpundParametricKnapsack} derived a PTAS for the general problem and, for the case that $\lambda \geq 0$ and $a_i,b_i \geq 0$ for all $i \in \{1,\ldots,n\}$, an FPTAS with a weakly polynomial running-time. Recently, \citet{ParametricKnapsackFPTAS} presented an FPTAS for the problem that works for arbitrary integral values of $a_i,b_i$ and runs in strongly polynomial time.

To the best of our knowledge, the parametric weight knapsack problem considered here was mentioned in only one publication so far. \citet{BurkardPferschyInverseParametricKnapsack} show that the optimal profit of the parametric weight knapsack problem can attain an exponential number of values in general, yielding that any exact algorithm for the problem must output an exponential number of knapsack solutions. The authors present a pseudo-polynomial algorithm for the \emph{inverse} problem, in which the largest value for the parameter is searched in order to achieve some given profit.

\subsection{Our contribution}

We present the first FPTAS for the parametric weight knapsack problem. In fact, this is the first approximation algorithm for the problem and even the first algorithmic approach at all. Our algorithm is based on two different ideas: The first approach simulates the well-known FPTAS for the traditional knapsack problem, in which the profits are scaled to a polynomial size. The second approach is based on an implicit scaling technique due to \citet{ErlebachMultiObjectiveKnapsack}, in which a $(1+\varepsilon)$-grid is laid over the search space of the underlying dynamic programming scheme such that only a (weakly) polynomial number of entries must be evaluated. Combining both approaches, the resulting FPTAS achieves a running time of
\begin{align*}
	&\OO\left( \frac{n^3}{\varepsilon^2} \cdot \min\left\{ \log^2 P, n^2 \right\} \cdot \right. \\
	&\hspace{6mm}\left.\min\left\{\log M, n \log \frac{n}{\varepsilon} \ \middle/\  \log\left(n \log \frac{n}{\varepsilon} \right) \right\} \right),
\end{align*}
which allows both a strongly polynomial implementation and three possibly more efficient weakly polynomial implementations.

\subsection{Organization}

The results of this paper are divided into three main parts. In Section~\ref{sec:TwoApprox}, we show how we can generalize the well-known $\frac{1}{2}$-approximation algorithm for the traditional knapsack problem to the case of parametric weights. This will build the foundation for the parametric FPTAS, which will be presented in Section~\ref{sec:FPTAS}. We will first recapitulate a basic FPTAS for the traditional knapsack problem in Section~\ref{sec:FPTAS:Traditional} and then extend it to the parametric case in Sections~\ref{sec:FPTAS:Parametric} and \ref{sec:FPTAS:Intervals}. Finally, we present an alternative approach that yields a weakly polynomial-time FPTAS in Section~\ref{sec:FPTAS2}. 

\section{Obtaining a parametric $\frac{1}{2}$-approx\-imation}
\label{sec:TwoApprox}

Our main FPTAS for the parametric knapsack problem relies on the well-known $\frac{1}{2}$-approximation algorithm for the traditional knapsack problem. In the non-parametric case, this algorithm proceeds as follows: In a first step, the algorithm sorts the items in increasing order of their ratios~$\frac{w_i}{p_i}$.\footnote{In its standard definition, the algorithm uses an decreasing order of the ratios~$\frac{p_i}{w_i}$. However, we will use the (equivalent) inverted formulation here since the parametric ratios will then take on affine forms.} The algorithm then constructs an intermediate solution~$x'$ in which $x'_i = 1$ if and only if $\sum_{j=1}^i w_j \leq W$ for each $i \in \{1,\ldots,n\}$. The algorithm then either returns~$x'$ or the solution which contains only an item with the largest profit among all items~$i$ with $w_i \leq W$. The resulting profit is denoted by $p^A$ and fulfills~$\frac{1}{2} \cdot p^* \leq p^A \leq p^*$, where $p^*$ is the optimal profit \citep{Knapsack}.

Now, in the parametric setting, the ratios~$\frac{w_i}{p_i}$ are affine functions of the form
\[ f_i(\lambda) \colonequals \frac{w_i(\lambda)}{p_i} = \frac{a_i + \lambda \cdot b_i}{p_i} = \frac{a_i}{p_i} + \lambda \cdot \frac{b_i}{p_i} \]
and the profit~$p^A$ of the approximation becomes a step function depending on $\lambda$. In our FPTAS, we will need a constant behavior of $p^A$, so we must divide the real line into intervals~$I_1,\ldots,I_q$ such that, for each $j \in \{1,\ldots,q\}$, it holds that $p^A(\lambda)$ is constant for $\lambda \in I_j$. A trivial bound for the number of necessary intervals is given by $q \in \OO(n^3)$ since the ordering can change up to $\OO(n^2)$~times (whenever two of the affine functions intersect) and since, for some given ordering, the number of elements that fit into the knapsack in the intermediate solution can change from $1$ to $n$ or vice versa in the worst case.

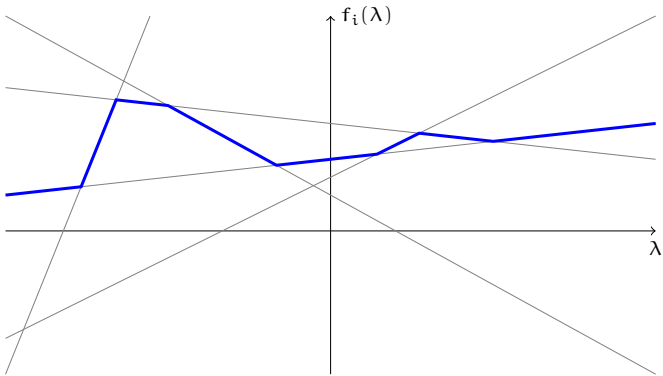
\begin{figure}[ht!]
	\begin{tikzpicture} \node[scale=0.95] {\begin{tikzpicture}
		\draw[->] (-4.5,0) -- (4.5,0) node[below] {\scriptsize $\lambda$};
		\draw[->] (0,-2) -- (0,3) node[right] {\scriptsize $f_i(\lambda)$};

		\draw[black!50!white,name path=F1] (-4.5,-2) -- (-2.5,3);
		\draw[black!50!white,name path=F2] (-4.5,0.5) -- (4.5,1.5);
		\draw[black!50!white,name path=F3] (-4.5,-1.5) -- (4.5,3);
		\draw[black!50!white,name path=F4] (-4.5,2.0) -- (4.5,1.0);
		\draw[black!50!white,name path=F5] (-4.5,3) -- (4.5,-2);

		\path[name intersections={of=F2 and F1,by=P1}];
		\path[name intersections={of=F1 and F4,by=P2}];
		\path[name intersections={of=F4 and F5,by=P3}];
		\path[name intersections={of=F5 and F2,by=P4}];
		\path[name intersections={of=F2 and F3,by=P5}];
		\path[name intersections={of=F3 and F4,by=P6}];
		\path[name intersections={of=F4 and F2,by=P7}];

		\draw[very thick,blue] (-4.5,0.5) -- (P1) -- (P2) -- (P3) -- (P4) -- (P5) -- (P6) -- (P7) -- (4.5,1.5);
	\end{tikzpicture}}; \end{tikzpicture}
	\caption{Plotting the functions~$f_i$ in the plane. The $k=3$ level is highlighted by the thick blue polygonal chain.}
	\label{fig:LevelFunction}
\end{figure}

This bound can be improved to $\OO(n^2)$, which can best be seen by considering so-called \emph{$k$-levels}: Let $S \colonequals \{f_i : i \in \{1,\ldots,n\}\}$ denote the set of all ratio functions as introduced above and, for some fixed $k \in \{1,\ldots,n\}$, let $g_k \colon \mathbb{R} \rightarrow \mathbb{R}$ denote the function mapping the parameter~$\lambda$ to the value of the $k$-th smallest function in $S$ at $\lambda$. The function~$g$ is called the \emph{$k$-level} and is both piecewise linear and continuous (see Figure~\ref{fig:LevelFunction}). Although a $k$-level can have super-linear many breakpoints in general, the total number of breakpoints among all levels is bounded by $\OO(n^2)$ (which is moreover clear since there are only $\OO(n^2)$ intersection points of the functions in $S$) \citep{EverettKLevel}.

For some given value of $\lambda$ and for some specific value of $k$, the intermediate solution~$x'$ as described above consists of the first $k$ items based on the current item ordering and can be associated with the $k$-level in a plane as depicted in Figure~\ref{fig:LevelFunction}. While $\lambda$ increases, the solution~$x'$ may change due to two reasons: Either, we enter a new line segment of the $k$-level\footnote{Actually, only an ``upward shifted corner'' of the $k$-level will generate a new solutions~$x'$ since only in this case the $k$-th smallest function leaves the knapsack and the composition of the knapsack changes.}, in which case the $k$-th smallest function and the greedy knapsack solution change; or we change the level, either since the first $k$ items do no longer fit into the knapsack or since now $k+1$ or more items fit. Both cases occur only $\OO(n^2)$~times since each of the knapsack solutions that corresponds to a linear line segment becomes feasible or infeasible at most once.

In summary, we get a partition of the real line into $h \in \OO(n^2)$ intervals~$I'_1,\ldots,I'_h$ in which the solution~$x'$ is constant. In addition, the most profitable item that fits into the knapsack can change up to $\OO(n)$~times. This yields $q \in \OO(n^2)$~final intervals~$I_1,\ldots,I_q$, which will be used in the FPTAS in Section~\ref{sec:FPTAS}. It is easy to see that these intervals can be constructed in $\OO(n^2 \log n)$~time.

\section{Obtaining a parametric FPTAS}
\label{sec:FPTAS}

Before we explain the parametric FPTAS in detail, we first recapitulate the basic FPTAS for the traditional (non-parametric) knapsack problem as introduced by \citet{LawlerKnapsack} since its way of proceeding is crucial for the understanding of the parametric version.

\subsection{Traditional FPTAS}
\label{sec:FPTAS:Traditional}

Consider the case of some fixed value for $\lambda$ such that the weights have a constant (and possibly negative) value~$w_i$. The basic FPTAS for the traditional knapsack problem is based on a well-known dynamic programming scheme, which was originally designed to solve the problem exactly in pseudo-polynomial time: Recall that $P$ denotes an upper bound on the maximum profit of a solution to the given instance. For $k \in \{0,\ldots,n\}$ and $p \in \{0,\ldots,P\}$, let $w(k,p)$ denote the minimum weight that is necessary in order to obtain a profit of exactly $p$ with the first $k$~items. For $k=0$, we set $w(0,p) = 0$ for $p=0$ and $w(0,p) = W+1$ for $p > 0$. For $k \in \{1,\ldots,n\}$ and for the case that $p_k \leq p$, we compute the values~$w(k,p)$ recursively by
\begin{align}\label{eqn:Recursion}
	w(k,p) = \min\{w(k-1,p), w(k-1,p-p_k) + w_k\},
\end{align}
representing the choice to either not pack the item or to pack it, respectively. Else, if $p_k > p$, we set $w(k,p) = w(k-1,p)$. The largest value of $p$ such that $w(n,p) \leq W$ then reveals the optimal solution to the problem. The procedure runs in pseudo-polynomial time $\mathcal{O}(nP)$.

The idea of the basic FPTAS is to scale down the item profits~$p_i$ to new values~$\pt_i \colonequals \floor{\frac{n \cdot p_i}{\varepsilon \cdot p^A}}$, where $p^A$ denotes the profit of the $\frac{1}{2}$-approximate solution described in Section~\ref{sec:TwoApprox}. The procedure runs in polynomial time~$\OO(\frac{n^2}{\varepsilon})$ since the maximum possible profit is now bounded by $\widetilde{P} \colonequals \frac{2n}{\varepsilon}$. The crucial observation is that we only lose a factor of $(1-\varepsilon)$ by scaling down the profits, so the solution obtained by the above dynamic programming scheme applied to the scaled profits yields a $(1-\varepsilon)$-approximate solution for the problem. We refer to \citep{LawlerKnapsack,Knapsack} for further details on the algorithm.

\subsection{Parametric FPTAS}
\label{sec:FPTAS:Parametric}

Now consider the parametric problem setting. Obviously, the weights of the items change as the parameter~$\lambda$ increases, but also the scaled profits may change with $\lambda$ since they depend on the profit~$p^A$ of the $\frac{1}{2}$-approximate solution. We partition the real line into the same intervals~$I_1,\ldots,I_q$ as described in Section~\ref{sec:TwoApprox} such that $p^A$ and, thus, the scaled profits are constant within each interval, yielding $q \in \OO(n^2)$~subproblems. In the following, we restrict our considerations to one specific such interval, so we may assume that the values~$\pt_i$ are constant for all considered values of $\lambda$.

As noted above, the item weights now depend on $\lambda$ such that new solutions may become feasible and current solutions may become infeasible as the parameter~$\lambda$ increases. The idea of the FPTAS is to consider each possible profit~$p \in \{0,\ldots,\widetilde{P}\}$ individually and to determine the values of $\lambda$ for which a profit of $p$ can be achieved by a feasible solution to the scaled knapsack instance.

\begin{figure}
	\begin{tikzpicture}[>=stealth',->]
		\tikzset{ns/.style={circle, draw=black, minimum size=2*(#1-\pgflinewidth), inner sep=0pt, outer sep=0pt}, ns/.default={9pt}}


		\node[ns] (V01) at (0,0) {\scriptsize $v_{0,0}$};
		\node[ns] (V11) at (0,-1.5) {\scriptsize $v_{1,0}$};
		\node[ns] (V21) at (0,-3.0) {\scriptsize $v_{2,0}$};
		\node[ns] (V31) at (0,-4.5) {\scriptsize $v_{3,0}$};
		\node at (0,-6.0) {\scriptsize $\vdots$};
		\node[ns] (VP1) at (0,-7.5) {\scriptsize $v_{\widetilde{P},0}$};

		\node[ns] (V02) at (2.5,0) {\scriptsize $v_{0,1}$};
		\node[ns] (V12) at (2.5,-1.5) {\scriptsize $v_{1,1}$};
		\node[ns] (V22) at (2.5,-3.0) {\scriptsize $v_{2,1}$};
		\node[ns] (V32) at (2.5,-4.5) {\scriptsize $v_{3,1}$};
		\node at (2.5,-6.0) {\scriptsize $\vdots$};
		\node[ns] (VP2) at (2.5,-7.5) {\scriptsize $v_{\widetilde{P},1}$};

		\node[ns] (V03) at (5,0) {\scriptsize $v_{0,2}$};
		\node[ns] (V13) at (5,-1.5) {\scriptsize $v_{1,2}$};
		\node[ns] (V23) at (5,-3.0) {\scriptsize $v_{2,2}$};
		\node[ns] (V33) at (5,-4.5) {\scriptsize $v_{3,2}$};
		\node at (5,-6.0) {\scriptsize $\vdots$};
		\node[ns] (VP3) at (5,-7.5) {\scriptsize $v_{\widetilde{P},2}$};

		\node at (6.5,0) {$\dots$};
		\node at (6.5,-1.5) {$\dots$};
		\node at (6.5,-3) {$\dots$};
		\node at (6.5,-4.5) {$\dots$};
		\node at (6.5,-6) {$\ddots$};
		\node at (6.5,-7.5) {$\dots$};

		\node[ns] (V0N) at (8,0) {\scriptsize $v_{0,n}$};
		\node[ns] (V1N) at (8,-1.5) {\scriptsize $v_{1,n}$};
		\node[ns] (V2N) at (8,-3.0) {\scriptsize $v_{2,n}$};
		\node[ns] (V3N) at (8,-4.5) {\scriptsize $v_{3,n}$};
		\node at (8,-6.0) {\scriptsize $\vdots$};
		\node[ns] (VPN) at (8,-7.5) {\scriptsize $v_{\widetilde{P},n}$};

		\path (V01) edge node[above] {\scriptsize $0$} (V02) (V01) edge node[right,pos=0.18] {\scriptsize $w_1(\lambda)$} (V22) (V02) edge node[above] {\scriptsize $0$} (V03) (V02) edge node[right,pos=0.4] {\scriptsize $w_2(\lambda)$} (V13) (V22) edge (V23); 
		\path (V11) edge (V12) (V11) edge (V32) (V12) edge (V13) (V12) edge (V23); 
		\path (V21) edge (V22) (V22) edge (V33) (V21) edge[path fading=east] (2,-5.25);
		\path (V31) edge (V32) (V31) edge[path fading=east] (2,-6.75) (V32) edge (V33) (V32) edge[path fading=east] (4.5,-5.625);
		\path (VP1) edge (VP2) (VP2) edge (VP3) (2.5,-6.375) edge[path fading=west] (VP3);
	\end{tikzpicture}
	\caption{The interpretation of the dynamic programming scheme as a shortest path problem. In this example, it holds that $\pt_1 = 2$ and $\pt_2 = 1$.}
	\label{fig:ShortestPath}
\end{figure}
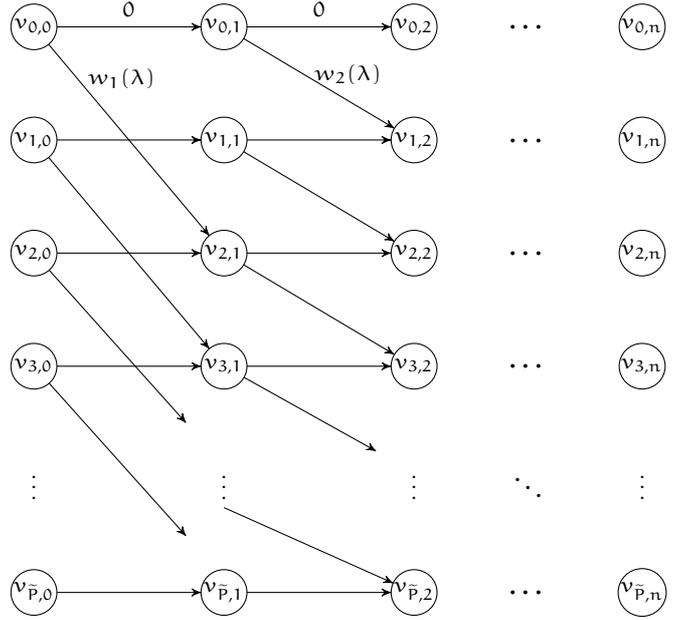

In the following, it will be useful to interpret the underlying dynamic programing scheme as a shortest path problem: Consider the acyclic graph shown in Figure~\ref{fig:ShortestPath}. For each $k \in \{0,\ldots,n\}$ and profit~$p \in \{0,\ldots,\widetilde{P}\}$, we insert a node~$v_{p,k}$ and connect it with $v_{p,k+1}$ via an edge with zero length (if $k \leq n-1$) as well as with $v_{p+\pt_{k+1},k+1}$ via an edge with length $w_{k+1}(\lambda)$ (if $k \leq n-1$ and $p + \pt_k \leq \widetilde{P}$). It is easy to see that the structure of the graph reflects the recursion given in \eqref{eqn:Recursion}, so a solution to the (scaled) knapsack instance then corresponds to the largest profit~$p$ such that the length of a shortest path from $v_{0,0}$ to $v_{p,n}$ is not larger than $W$.

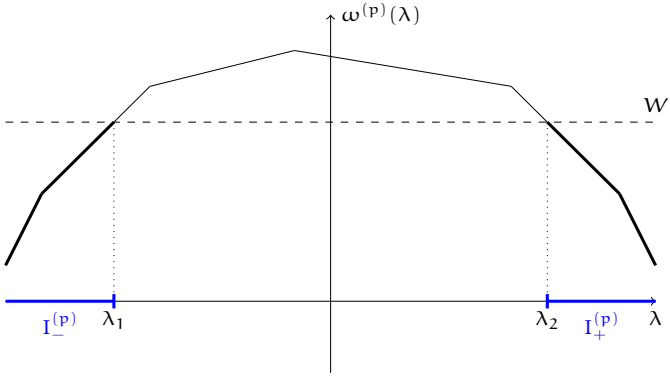
\begin{figure}[ht!]
	\begin{tikzpicture} \node[scale=0.95] {\begin{tikzpicture}
		\draw[->] (-4.5,-1) -- (4.5,-1) node[below] {\scriptsize $\lambda$};
		\draw[->] (0,-2) -- (0,3) node[right] {\scriptsize $\omega^{(p)}(\lambda)$};

		\draw[dashed] (-4.5,1.5) -- (4.5,1.5) node[above] {\scriptsize $W$};

		\draw[very thick] (-4.5,-0.5) -- (-4.0,0.5);
		\draw[very thick] (-4.0,0.5) -- (-3.0,1.5);
		\draw[thin] (-3.0,1.5) -- (-2.5,2.0);
		\draw[thin] (-2.5,2.0) -- (-0.5,2.5);
		\draw[thin] (-0.5,2.5) -- (2.5,2.0);
		\draw[thin] (2.5,2.0) -- (3.0,1.5);
		\draw[very thick] (3.0,1.5) -- (4.0,0.5);
		\draw[very thick] (4.0,0.5) -- (4.5,-0.5);

		\draw[dotted] (-3.0,1.5) -- (-3.0,-1) node[below] {\scriptsize $\lambda_1$};
		\draw[dotted] (3.0,1.5) -- (3.0,-1) node[below] {\scriptsize $\lambda_2$};

		\draw[blue,very thick] (-4.5,-1) -- (-3.0,-1) node[below,pos=0.5] {\scriptsize $I_-^{(p)}$};
		\draw[blue,very thick] (3.0,-1) -- (4.5,-1) node[below,pos=0.5] {\scriptsize $I_+^{(p)}$};
		\draw[blue,very thick] (-3.0,-0.9) -- (-3.0,-1.1);
		\draw[blue,very thick] (3.0,-0.9) -- (3.0,-1.1);
	\end{tikzpicture}}; \end{tikzpicture}
	\caption{The function~$\omega^{(p)}$ for some specific profit~$p$ (black) together with the feasibility sets~$I^{(p)}_-$ and $I^{(p)}_+$ (blue).}
	\label{fig:FeasibilityIntervals}
\end{figure}

Now consider some specific profit~$p \in \{0,\ldots,\widetilde{P}\}$. In general, there may be a super-polynomial number of paths in the graph that lead from $v_{0,0}$ to $v_{p,n}$. The lengths of these paths are described by affine functions depending on the parameter~$\lambda$ and may increase or decrease as $\lambda$ increases. Since we are interested in shortest paths, we can restrict our considerations to the function~$\omega^{(p)} \colon \mathbb{R} \rightarrow \mathbb{R}$ mapping the parameter~$\lambda$ to the length of a shortest path from $v_{0,0}$ to $v_{p,n}$. Clearly, the function $\omega^{(p)}$ is concave, continuous, and piecewise linear since it is the point-wise minimum of finitely many affine functions (see Figure~\ref{fig:FeasibilityIntervals}). Whenever $\omega^{(p)}(\lambda) \leq W$, the knapsack solution induced by a shortest path from $v_{0,0}$ to $v_{p,n}$ is feasible and attains a profit of exactly~$p$. Besides the special cases that, for all $\lambda \in \mathbb{R}$, $\omega^{(p)}(\lambda) \leq W$ or, for all $\lambda \in \mathbb{R}$, $\omega^{(p)} > W$, there is at most one interval~$I_-^{(p)} \colonequals (-\infty,\lambda_1]$ and at most one interval~$I_+^{(p)} \colonequals [\lambda_2,+\infty)$ with $\omega^{(p)}(\lambda_1) = \omega^{(p)}(\lambda_2) = W$ containing the values of $\lambda$ for which $\omega^{(p)}(\lambda) \leq W$ (see Figure~\ref{fig:FeasibilityIntervals}). We call these intervals the \emph{feasibility intervals of $p$} in the following. We associate with each feasibility interval exactly one knapsack solution, which remains feasible throughout the whole interval, by selecting the solution which corresponds to the intersection at $\lambda_1$ and $\lambda_2$, respectively. Note that this solution may not give a shortest path for the \emph{whole} interval but the corresponding knapsack solution remains feasible and attains the desired profit~$p$.

In Section~\ref{sec:FPTAS:Intervals}, we show that we can detect these intervals in polynomial time for some specific profit~$p$. Performing these steps for all attainable profits in $\{0,\ldots,\widetilde{P}\}$, we obtain at most $2 \cdot \frac{2n}{\varepsilon}$ many knapsack solutions together with their feasibility intervals. By reevaluating the maximum attainable profit at each boundary value of these intervals, we obtain parametric solutions for the scaled knapsack instance and, thus, an FPTAS for the parametric weight knapsack problem.

\subsection{Computing the feasibility intervals}
\label{sec:FPTAS:Intervals}

It remains to show how we can determine the feasibility intervals efficiently. For some specific profit~$p$, we can identify the two intervals (or detect the special cases that all points are feasible or that there are no feasible points at all) by two different methods.

Both methods rely on the same observation: For some given candidate value~$\lambda \in \mathbb{R}$, we can determine in $\OO(\frac{n^2}{\varepsilon})$~time if it is smaller, larger, or equal to the value~$\lambda_1$ (if such a value exists) as introduced above. This observation is based on the fact that a value of $\lambda$ is too small if and only if the length~$\omega^{(p)}(\lambda) \colonequals \alpha + \lambda \cdot \beta$ of a shortest path is strictly smaller than $W$ \emph{and} the slope~$\beta$ is positive (if the slope is zero, we have the special case that $\omega^{(p)}(\lambda) < W$ for all $\lambda \in \mathbb{R}$ due to the concavity of $\omega^{(p)}$). Conversely, a given value for $\lambda$ is too large if $\omega^{(p)} > W$ \emph{or} the slope $\beta$ is negative. It then holds that $\lambda = \lambda_1$ if $\omega^{(p)} = W$ and $\beta > 0$. Analogous arguments apply to the case of $\lambda_2$.

This observation can be incorporated into a binary search in order find the value of $\lambda_1$ if it exists (the case of $\lambda_2$ works analogously). Consider two $v_{0,0}$-$v_{p,n}$-paths~$P_1$ and $P_2$ with lengths~$w_1(\lambda) \colonequals \alpha_1 + \lambda \cdot \beta_1$ and $w_2(\lambda) \colonequals \alpha_2 + \lambda \cdot \beta_2$ and assume that $\beta_1 \neq 0$ and $\beta_2 \neq 0$. The critical values of $\lambda$ at which the paths become (in)feasible are given by~$\frac{W - \alpha_1}{\beta_1}$ and $\frac{W - \alpha_2}{\beta_2}$, respectively, and are thus contained in the interval~$[-(n+1)M;(n+1)M]$. Moreover, in case that the two critical values are not equal, they differ by at least~$\frac{1}{n^2 M^2}$ since
\begin{align*}
	& \left| \frac{W - \alpha_1}{\beta_1} - \frac{W - \alpha_2}{\beta_2} \right| \\
	=& \left| \frac{(W - \alpha_1) \cdot \beta_2  - (W - \alpha_2) \cdot \beta_1}{\beta_1 \cdot \beta_2} \right| \geq \frac{1}{n^2 M^2}.
\end{align*}
Hence, we only need to scan the interval~$[-(n+1)M;(n+1)M]$ in steps of length $\delta \colonequals \frac{1}{2} \cdot \frac{1}{n^2 M^2}$: Consider a path~$P_1$ in the set of the desired shortest paths determining the value of $\lambda_1$. Every other path with positive slope either has the same length as $P_1$ at $\lambda_1$ or attains the value~$W$ at some point that is strictly less than $\lambda_1 - \delta$. Hence, whenever we encounter a situation in which
\[ \omega^{(p)}(j \cdot \delta) \leq W \ \ \text{ and }\ \ \omega^{(p)}((j+1) \cdot \delta) \geq W \]
for some $j \in \mathbb{Z}$, we know that the shortest path that is obtained at $j \cdot \delta$ is also a shortest path at $\lambda_1$. In this case, we can determine the feasibility interval~$I_-^{(p)}$ exactly and assign it with the corresponding knapsack solution. Moreover, we do not need to scan these points sequentially, but are able to perform a binary search with a logarithmic number of steps. In combination with the running time of $\OO(\frac{n^2}{\varepsilon})$ to compute the shortest paths, this approach yields a weakly polynomial overhead of
\begin{align*}
	& \OO\left(\frac{n^2}{\varepsilon} \cdot \log (nM \cdot n^2 M^2) \right) = \OO\left(\frac{n^2}{\varepsilon} \cdot \log M \right),
\end{align*}
where the last equality follows from the fact that $M \geq n$ in Equation~\eqref{eqn:BigM}.

As a second approach, we can perform two of \citets{MegiddoCombinatorialOptimization} parametric searches in order to determine the values of $\lambda_1$ and $\lambda_2$: In contrast to the first approach, we treat the variable~$\lambda$ as a \emph{symbolic} variable and simulate the shortest path computation step by step until we need to stop in order to resolve a comparison of two affine functions, whose intersection yields a candidate value~$\lambda$. For each such candidate value, we can proceed as above in order to determine if it is too small, too large, or equal to $\lambda_1$ (or $\lambda_2$, respectively). This would yield an overhead of $\OO\left((\frac{n^2}{\varepsilon})^2 \right)$ since we need to recompute a shortest path in each step of the simulation in the worst case. Using techniques described in \citep{BurkardPferschyInverseParametricKnapsack}, this running time can be significantly improved to
\begin{align*}
 	\OO\left( \frac{n^3}{\varepsilon} \cdot \log \frac{n}{\varepsilon} \ \middle/\  \log\left(n \log \frac{n}{\varepsilon} \right) \right).
\end{align*}
We refer to \citep{MegiddoCombinatorialOptimization,MegiddoParallel} for further details on the parametric search technique and to \citep{BurkardPferschyInverseParametricKnapsack} for its application to knapsack problems.

In summary, for some profit~$p \in \{0,\ldots,\widetilde{P}\}$, we can determine the feasibility intervals in
\begin{align*}
 	\OO\left( \frac{n^2}{\varepsilon} \cdot \min\left\{ \log M, n \log \frac{n}{\varepsilon} \ \middle/\  \log\left(n \log \frac{n}{\varepsilon} \right) \right\} \right)
\end{align*}
time. Since we need to repeat these steps for $\OO(\frac{n}{\varepsilon})$ possible profits and for $\OO(n^2)$~intervals, we get one of the main results of this paper:
\begin{theorem}\label{thm:FPTAS1}
	There is an FPTAS for the parametric weight knapsack problem running in total time
	\begin{align*}
		\OO\left( \frac{n^5}{\varepsilon^2} \cdot \min\left\{ \log M, n \log \frac{n}{\varepsilon} \ \middle/\  \log\left(n \log \frac{n}{\varepsilon} \right) \right\} \right).
	\end{align*}\qed
\end{theorem}

\section{An alternative Approach}
\label{sec:FPTAS2}

A significant contribution to the total running time of the approach discussed in Section~\ref{sec:FPTAS} comes with the explicit scaling of the profits: We needed to divide the problem into $\OO(n^2)$ subproblems in order to guarantee constant values of the scaled profits. In this section, we discuss a solution that does not scale the profits explicitly, yielding a second FPTAS with a weakly polynomial running time. 

Our second approach relies on an implicit scaling technique introduced by \citet{ErlebachMultiObjectiveKnapsack}, originally designed for the multi-objective knapsack problem. In their FPTAS, the authors stick to the original (unscaled) profits, but perform the dynamic programming scheme only for a polynomial number of profits: Instead of considering each integral profit, the profit space is being reduced to the points in the set
\[ S \colonequals \left\{ (1 + \varepsilon)^{\frac{i}{n}} : i \in \{0,\ldots,\ceil{n \log_{1 + \varepsilon} P}\} \right\}. \]
The entries of $w(\cdot,\cdot)$ are only computed for the points in the set~$S$ such that the dynamic programming scheme again runs in polynomial time. In order for the recursion to be well-defined, the term $p - p_k$ is ``rounded up'' to the next value in $S$ in Equation~\eqref{eqn:Recursion}. Hence, $w(k,p)$ denotes the minimum weight that is necessary to achieve a profit of \emph{at least} $p$. In can be seen that, in the case of constant weights, this approach yields an FPTAS for the knapsack problem (cf. \citep{ErlebachMultiObjectiveKnapsack}).

Analogously to the FPTAS derived in Section~\ref{sec:FPTAS}, the recursive formulae~\eqref{eqn:Recursion} to compute the approximate solution induce an interpretation as a shortest path problem in an acyclic graph. Consequently, the tools derived in Section~\ref{sec:FPTAS} apply to this graph structure: We can determine the feasibility intervals for each attainable profit (which is now contained in the set~$S$) subsequently and use them in order to obtain a partition of the real line into different $(1-\varepsilon)$-approximate solutions. This graph still has $\OO(n)$~vertical layers, but the number of nodes per layer increases from $\OO(\frac{n}{\varepsilon})$ to
\begin{align*}
	\OO\left( \ceil{n \log_{1 + \varepsilon} P} \right) &= \OO\left( n \cdot \frac{\log P}{\log (1 + \varepsilon)} \right) \\
	& = \OO\left( \frac{n}{\varepsilon} \cdot \log P \right).
\end{align*}
Hence, since the complexities to solve the dynamic programming scheme and to iterate through each possible profit both increase by $\OO(\log P)$ likewise, we get an additional overhead of $\OO(\log^2 P)$ that comes with this alternative approach \emph{but} do no longer need to divide the problem into $\OO(n^2)$~subproblems. This yields the second main theorem of this paper:
~\\
~\\
\begin{theorem}\label{thm:FPTAS2}
	There is an FPTAS for the parametric weight knapsack problem running in total time
	\begin{align*}
		&\OO\left( \frac{n^3}{\varepsilon^2} \cdot \log^2 P \cdot \right. \\
		&\hspace{6mm}\left.\min\left\{ \log M, n \log \frac{n}{\varepsilon} \ \middle/\  \log\left(n \log \frac{n}{\varepsilon} \right) \right\} \right). \tag*{\qed}
	\end{align*}
\end{theorem}
Combining the results of Theorem~\ref{thm:FPTAS1} and Theorem~\ref{thm:FPTAS2}, we get the following final result for the approximability of the parametric weight knapsack problem:
\begin{theorem}\label{thm:FPTAS}
	There is an FPTAS for the parametric weight knapsack problem running in total time
	\begin{align*}
		&\OO\left( \frac{n^3}{\varepsilon^2} \cdot \min\left\{ \log^2 P, n^2 \right\} \cdot \right. \\
		&\hspace{6mm}\left.\min\left\{\log M, n \log \frac{n}{\varepsilon} \ \middle/\  \log\left(n \log \frac{n}{\varepsilon} \right) \right\} \right). \tag*{\qed}
	\end{align*}
\end{theorem}

\bibliographystyle{elsarticle-num-names}
\bibliography{/Users/holzhaus/Documents/Forschung/literature}

\end{document}